\documentstyle[multicol,aps,epsf]{revtex}
\begin{document}
\title{Dynamic Crossover to Tricriticality and Anomalous Slowdown of Critical
Fluctuations by Entanglements in Polymer Solutions}
\author{A. F. Kostko, M. A. Anisimov, and J. V. Sengers}
\address{Institute for Physical Science and Technology and Department of Chemical\\
Engineering \\
University of Maryland, College Park, MD 20742}
\date{\today}
\maketitle

\begin{abstract}
We have performed accurate dynamic light-scattering measurements near
critical demixing points of solutions of polystyrene in cyclohexane with
polymer molecular weight ranging from 200,000 to 11.4 million. Two dynamic
modes have been observed, ''slow'' and ''fast'', that result from a coupling
between diffusive relaxation of critical fluctuations of the concentration
and visco-elastic relaxation associated with the entanglement network of the
polymer chains. The coupling with the visco-elastic mode causes an
additional slowdown of the critical mode on top of the uncoupled diffusion
mode. By implementing crossover from critical to {\it theta}-point
tricritical behavior for both static and dynamic properties, we are able to
present a quantitative description of the phenomenon and to obtain a scaling
of the visco-elastic parameters as a function of the molecular weight.
\end{abstract}


\protect\tightenlines
\begin{multicols}{2}

Critical phenomena in high-molecular-weight polymer solutions differ from
critical phenomena in simple fluids. In polymer solutions, the thermodynamic
properties and the static correlations near the critical point of mixing are
determined by a competition of two mesoscale lengths, namely, the
correlation length of critical fluctuations of the concentration (tuned by
the distance to the critical point) and the radius of gyration of the
polymer molecules (tuned by the molecular weight) \cite{MelnAnis,PREinPress}%
. In the asymptotic vicinity of the critical point, the correlation length
becomes much larger than the radius of gyration and the polymer solution
exhibits Ising critical behavior. With increase of the polymer molecular
weight and, hence, of the radius of gyration, the range of asymptotic Ising
critical behavior shrinks, ultimately yielding to {\it theta}-point
tricritical behavior\cite{PREinPress}. This competition of two mesoscales
arises from a coupling between two different order parameters belonging to
two different static universality classes, namely, one associated with phase
separation and another one with self-avoiding-walk singularities of long
polymer chains \cite{deGennes}. Such a competition of two mesoscales and,
consequently, crossover from critical to multicritical behavior are expected
to be a general feature of phase transitions with coupled order parameters 
\cite{Domb}. Hence, it is natural to expect that the dynamic critical
behavior near the {\it theta} point will be affected by a coupling between
two soft dynamic modes associated with the two order parameters \cite{Tanaka}%
.

In the present communication we report a study of dynamic correlations in
near-critical solutions of polystyrene in cyclohexane at the critical volume
fraction $\phi _{{\rm c}}$ as a function of temperature and of polymer
molecular weight,\ ranging from 200,000 to 11.4 million. Specifically, we
have found that starting with a molecular weight of about one million, two
effective dynamic modes emerge, that result from a coupling between two soft
critical modes, a diffusion mode (association with the decay of critical
fluctuations) and a visco-elastic mode (associated with entanglements of
long polymer chains). Very close to the critical point, as expected, the
slow mode becomes critical. However, we have found that this mode being
trapped by the entanglements, may become an order slower as compared to
uncoupled critical slowdown. Moreover, the diffusivity, like static
properties \cite{PREinPress}, exhibits crossover from critical (Ising) to 
{\it theta}-point tricritical (mean-field) behavior, becoming almost
``classical'' at the highest molecular weight. In fact, the coupling of two
modes and the crossover to tricriticality are just two sides of the same
physical phenomenon, caused by an interaction of two order parameters
resulting in the emergence of a tricritical point.

Coupling between diffusion and entanglement has been discussed in the
literature for a long time. It was predicted by Brochard and De Gennes \cite
{BG77Brochard83} and first detected with dynamic light scattering in
non-critical polymer solutions by Adam and Delsanti \cite{AdamDelsanti} and
later by Jian {\it et al.} \cite{JianBrown96} and Nicolai {\it et al. }\cite
{NicolaiStepanek}. Ritzl{\it \ et al. }\cite{RitzlBelkoura} observed two
dynamic modes near the demixing critical temperature in a
polystyrene-cyclohexane solution with a polystyrene molecular weight $M_{%
{\rm w}}=0.96$ million. Most recently, Tanaka {\it et al}. \cite{Tanaka}.
observed two dynamic modes in near-critical solutions of
high-molecular-weight nearly monodisperse polystyrene in diethyl malonate
(explicitly shown for $M_{{\rm w}}=3.84$ million with a polydispersity index 
$M_{{\rm w}}/M_{{\rm n}}=1.04$, where subscripts ``w'' and ``n'' denote
weight and number averaging). Our experiments reveal how the coupled
critical dynamics in polymer solutions varies with increasing molecular
weight. Moreover, making use of theoretical predictions previously developed
for the static and dynamic crossover behavior from Ising to mean-field
(tricritical) behavior, we are able to present a quantitative description of
the phenomenon.

The original light-scattering setup as well as measurements for a
moderate-molecular-weight sample PS1 ($M_{{\rm w}}$=1.96$\cdot $10$^{5}$, $%
M_{{\rm w}}/M_{{\rm n}}=1.02$, $\phi _{{\rm c}}=0.066$, $T_{{\rm c}}=296.47$
K) have been described in a previous publication \cite{Jacob:01}. We have
measured the intensity of scattering and the dynamic correlation function
with a linear (PhotoCor-SP) \cite{PhotocorSP} and a logarithmic (ALV-5000/E)
correlator. The reduced distance of the temperature $T$ to the critical
temperature $T_{{\rm c}}$, $(T-T_{{\rm c}})/T,$ varied from 0.1 to 10$^{-5}$%
. Four high-molecular-weight samples were studied: PS2: $M_{{\rm w}}$=1.12$%
\cdot $10$^{6}$ ($M_{{\rm w}}/M_{{\rm n}}=1.06$, $\phi _{{\rm c}}=0.033$, $%
T_{{\rm c}}=303.09$ K), PS3: $M_{{\rm w}}$=1.95$\cdot $10$^{6}$ ($M_{{\rm w}%
}/M_{{\rm n}}=1.04$, $\phi _{{\rm c}}=0.024$, $T_{{\rm c}}=304.31$ K), PS4: $%
M_{{\rm w}}=3.9\cdot 10^{6}$ ($M_{{\rm w}}/M_{{\rm n}}=1.05$, $\phi _{{\rm c}%
}=0.018$, $T_{{\rm c}}=304.80$ K), and PS5: $M_{{\rm w}}=11.4\cdot 10^{6}$ ($%
M_{{\rm w}}/M_{{\rm n}}=1.09$, $\phi _{{\rm c}}=0.011$, $T_{{\rm c}}=305.95$
K).\ The radii of gyration $R_{{\rm g}}$ of the polymers are 12 nm for PS1
(from a SANS experiment \cite{Meln}), 28 nm, 37 nm, 52 nm, and 89 nm for
PS2--PS5, respectively (extrapolated as $R_{{\rm g}}\sim M_{{\rm w}}^{0.5}$%
). We used two scattering angles, $30^{\circ }$and $150^{\circ }$, which
correspond to the wave numbers $q=4\pi n/\lambda \sin ^{2}(\theta /2)$\ ($n$%
\ is the refractive index of the solution, $\lambda $ is the wave length of
light, and $\theta $ is the angle of scattering) equal to $7.27\cdot 10^{-3}$
nm$^{-1}$ and $2.73\cdot 10^{-2}$ nm$^{-1}$, respectively. We have performed
a Monte-Carlo simulation \cite{Malte} of the intensity of the multiple
scattering for our samples and found that at $(T-T_{{\rm c}})/T>10^{-5}$
double scattering accounts for almost the entire correction to static
properties. The double-scattering correction to the dynamic correlation
function at scattering angles of $30^{\circ }$and $150^{\circ }$ is expected
to be relatively small \cite{Ferrell} and, therefore,\ has been neglected.

\begin{figure*}
\centering
%\hspace{10pt}
\epsfbox{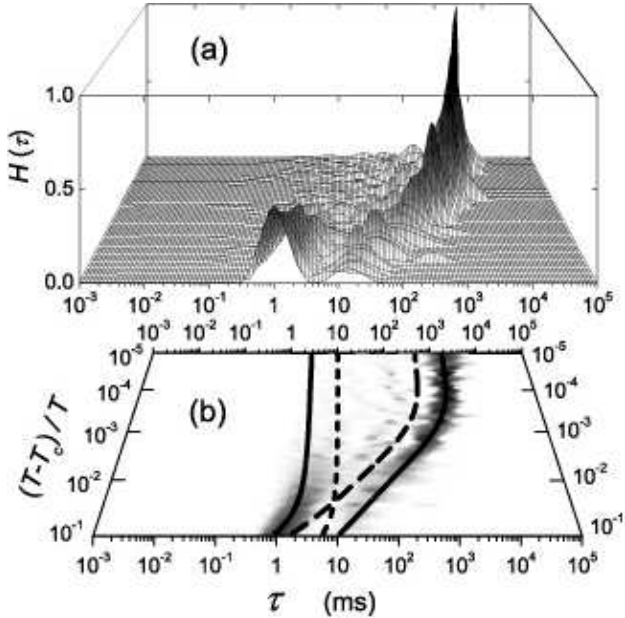}
%\vspace{0.4cm}
\begin{minipage}[]{8.5cm}
\protect\caption{
(a) Equal-area representation of the relaxation-time distribution
extracted from the intensity correlation function for the critical
polystyrene-cyclohexane solution PS5 ($M_{{\rm w}}=11.4\times 10^{6}$, $%
\theta =30^{\circ }$) as a function of the reduced temperature ($T-T_{{\rm c}%
}$)$/T$. (b) Projection of the relaxation-time distribution plot. The solid
curves represent the relaxation times $\tau _{-}$ and $\tau _{+}$ of the
effective modes as calculated from Eq. (2). The long-dashed curve represents
the pure critical diffusion time $\tau _{q}$ as calculated from Eq. (3). The
short-dashed curve represents the pure visco-elastic relaxation time $\tau _{%
{\rm ve}}$.
}
\end{minipage}
\label{f:1}
\end{figure*}

The correlation function for PS1 was found to be a single exponential and no
additional mode was detected \cite{Jacob:01}. For this sample the values of $%
qR_{{\rm g}}$ for the two scattering angles of $30^{\circ }$and $150^{\circ
} $ are 0.086 and 0.32, respectively. For all other samples with larger
polymer molecules we have found the dynamic correlation function to be not a
single exponential. The distributions $H(\tau )$ of decay-times $\tau $ have
been extracted from the correlation functions with the ALV-5000/E built-in
regularization procedure.

We show in Fig. 1a, as an example, a three-dimensional plot of this
distribution for the sample PS5 and the angle of $30^{\circ }$ ($qR_{{\rm g}%
}=0.65$) as a function of the distance to the critical temperature. All
distributions are normalized by their integrals, thus the narrower the
distribution, the higher the peak. One can clearly distinguish two modes
that change significantly upon approach to the critical temperature. Far
away from the critical temperature one can see a ``fast'' mode with a high
peak and a ``slow'' mode with a low peak. Upon approaching the critical
temperature, the intensity of the fast mode decreases and that of the slow
mode rapidly increases. The shape of the corresponding dynamic correlation
function in the near-vicinity of the critical point is again close to a
single exponential and the characteristic correlation time exceeds one
second. Two other examples, PS4 at $150^{\circ }$ ($qR_{{\rm g}}=1.42$) and
PS5 at $150^{\circ }$ ($qR_{{\rm g}}=2.43$), shown in Figs. 2a and 3a,
exhibit similar behavior with a more pronounced saturation of the intensity
of the slow mode for larger $qR_{{\rm g}}$.

%\vspace{-0.2cm}

\begin{figure*}
\centering
%\hspace{10pt}
\epsfbox{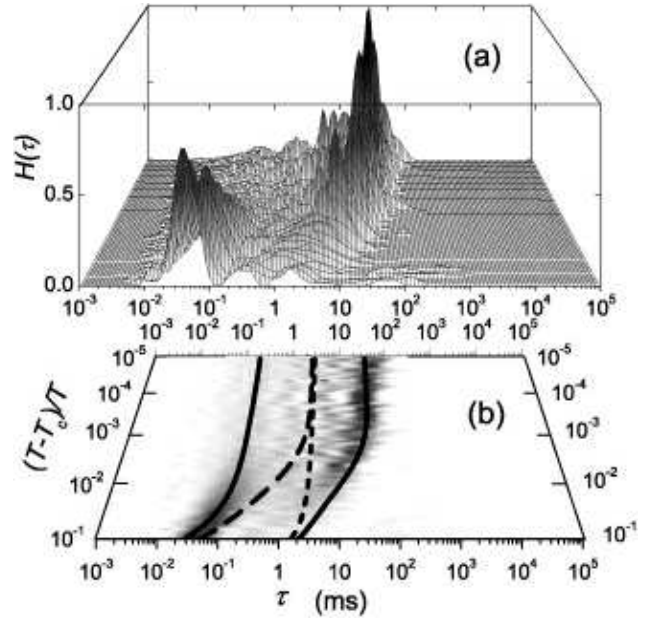}
\vspace{0.2cm}
\begin{minipage}[]{8.5cm}
\protect\caption{
Same as Fig. 1, but for PS4 ($M_{{\rm w}}=3.9\times 10^{6}$, $%
\theta =150^{\circ }$).
}
\end{minipage}
\label{f:1}
\end{figure*}

We have interpreted\ these results quantitatively in terms of a coupling
between two original soft modes, namely, a viscoelastic mode associated with
entanglements of polymer chains and the diffusion mode associated with the
relaxation of critical fluctuations. Neither of the two effective modes
actually observed is a pure viscoelastic or a diffusion mode. Instead, the
observed modes emerge as a result of the interaction of the original
uncoupled modes. Following Tanaka {\it et al}. \cite{Tanaka}, we use the
theory of Brochard and De Gennes \cite{BG77Brochard83} to represent the two
effective dynamic modes with corresponding characteristic times, ``slow'' $%
\tau _{-}$ and ``fast'' $\tau _{+}.$ Then the normalized time-dependent
intensity-correlation function, $g_{2}(t)$, can be approximated as

\begin{equation}
g_{2}(t)=1+\left[ f_{+}\exp \left( -\frac{t}{\tau _{+}}\right) +f_{-}\exp
\left( -\frac{t}{\tau _{-}}\right) \right] ^{2}  \label{g2}
\end{equation}
with decay times 
\begin{equation}
\frac{1}{\tau _{\pm }}=\frac{1+q^{2}\xi _{{\rm ve}}^{2}+\frac{\tau _{{\rm ve}%
}}{\tau _{q}}\pm \sqrt{\left( 1+q^{2}\xi _{{\rm ve}}^{2}+\frac{\tau _{{\rm ve%
}}}{\tau _{q}}\right) ^{2}-4\frac{\tau _{{\rm ve}}}{\tau _{q}}}}{2\tau _{%
{\rm ve}}}  \label{tau+-}
\end{equation}
and amplitudes $f_{\pm }=\pm \left[ \frac{\tau _{{\rm ve}}}{\tau _{\pm }}%
-\left( 1+q^{2}\xi _{{\rm ve}}^{2}\right) \right] \left( \frac{\tau _{{\rm ve%
}}}{\tau _{+}}-\frac{\tau _{{\rm ve}}}{\tau _{-}}\right) ^{-1}$.

In Eq. (\ref{tau+-}), $\tau _{{\rm ve}}$ is the $q$-independent
visco-elastic relaxation time determined by entanglements of polymer chains, 
$\tau _{q}$ is the $q$-dependent diffusion relaxation time of critical
concentration fluctuations, while $\xi _{{\rm ve}}$ is a mesoscopic
visco-elastic length scale \cite{DoiOnuki}. We have calculated the
temperature dependence of the critical diffusion mode with the help of
dynamic scaling theory \cite{Kawasaki,Burstyn} accounting for a
``background'' (classical) contribution $D_{0}q^{2}$ to the diffusion
coefficient and for a crossover behavior of the ``critical'' (scaling) term:

\begin{equation}
\tau _{q}^{-1}=\frac{k_{{\rm B}}T}{6\pi \eta \xi }K(x)\bigg[1+\bigg(\frac{x}{%
2}\bigg)^{2}\bigg]^{z_{\eta }/2}\Omega \left( q_{{\rm D}}\xi \right)
+D_{0}q^{2}\text{ .}  \label{tauq}
\end{equation}
In Eq.~(\ref{tauq}) $K(x)$ is %well known 
Kawasaki's function \cite{Kawasaki}: $%
K(x)=3/(4x^{2})\left[ 1+x^{2}+\left( x^{3}-x^{-1}\right) \arctan (x)\right] $%
; $\Omega \left( q_{{\rm D}}\xi \right) $ is a dynamic crossover function
approximated as \cite{Kiselev,LuettmerSengers}$\ \Omega \left( q_{{\rm D}%
}\xi \right) =2/\pi \arctan \left( q_{{\rm D}}\xi \right) $; and $D_{0}$ the
classical diffusion coefficient: $D_{0}=k_{{\rm B}}T\left[ 1+x^{2}\right]
(6\pi \eta _{{\rm b}}q_{{\rm D}}\xi ^{2})^{-1}$, where $k_{{\rm B}}$ is
Boltzmann's constant, $\eta $ is the viscosity of the solution, $\eta _{{\rm %
b}}$ the regular part (background) of the viscosity; $x=q\xi $ with $\xi $
being the static correlation length measured independently \cite{PREinPress}%
; $z_{\eta }=0.065$\ is a universal dynamic scaling exponent; $q_{{\rm D}}$
is a characteristic cutoff wavenumber expected to be inversely proportional
to the radius of gyration. It turned out that a good description of the
experimental data is obtained if we adopt $q_{{\rm D}}=\xi _{{\rm ve}}^{-1}$%
, which indeed scales approximately as $R_{{\rm g}}^{-1}$.

Asymptotically close to the critical temperature the classical term is
negligible, the crossover function $\Omega \left( q_{{\rm D}}\xi \right)
\rightarrow 1$, and $\tau _{q}$ is described by the standard theory
developed for simple fluids \cite{Kawasaki}. Incidentally, the static
correlation length obeys the Ising asymptotic power law: $\xi \sim (T-T_{%
{\rm c}})^{-0.63}$ \cite{PREinPress}. Further away from the critical
temperature or with increase in the molecular weight, {\it i.e.}, when $R_{%
{\rm g}}/\xi $ increases, $\Omega \left( q_{{\rm D}}\xi \right) $ vanishes
and $\tau _{q}$ is described by the classical diffusion term with the
classical correlation length $\xi \sim (T-T_{{\rm c}})^{-0.5}$. We have
found that accounting for crossover to tricriticality in Eq. (\ref{tauq}) is
necessary to obtain quantitative description of the critical dynamics near
the {\it theta} point. The shear viscosity, $\eta $, is expected to behave
as $\eta \simeq \eta _{{\rm 0}}\left[ \left( T-T_{{\rm c}}\right) /T\right]
^{-0.04}$ \cite{Kawasaki,Burstyn}. The coefficient $\eta _{{\rm 0}}(T)$ is
proportional to $\eta _{{\rm b}}(T)$ for which we assumed the same
Arhenius-type temperature dependence as that of the solvent, while treating $%
\eta _{{\rm 0}}(T=T_{{\rm c}})$ as an adjustable constant for each sample.
With the crossover critical diffusion mode calculated from Eq. (\ref{tauq}),
we have obtained two $q$-independent parameters, namely, the visco-elastic
time $\tau _{{\rm ve}}$ and the visco-elastic length $\xi _{{\rm ve}}$, by
fitting two (experimentally obtained) dynamic modes to Eq. (\ref{tau+-})
simultaneously for both scattering angles.

\begin{figure*}
\centering
%\hspace{10pt}
\epsfbox{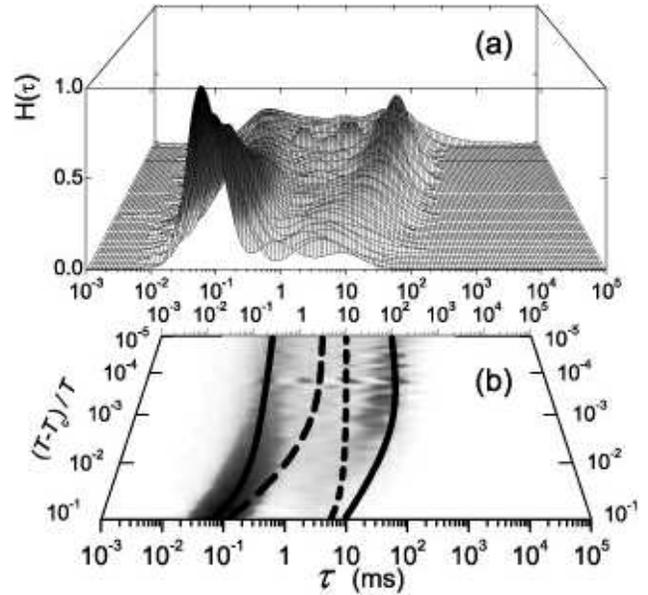}
\vspace{0.1cm}
\begin{minipage}[]{8.5cm}
\protect\caption{
Same as Fig. 1, but for PS5 ($M_{{\rm w}}=11.4\times 10^{6}$, $%
\theta =150^{\circ }$).
}
\end{minipage}
\label{f:1}
\end{figure*}

The resulting temperature dependences of the characteristic times are shown
on the projections of Figs. 1a-3a. (denoted as Figs. 1b-3b). As the
visco-elastic time is proportional to $\eta _{{\rm s}}/T$ ($\eta _{{\rm s}}$
is\ the solvent viscosity) \cite{BrochdeGenn82}, it exhibits a slight
temperature dependence. It is seen that neither the observed slow mode nor
the observed fast mode can be associated with a pure critical diffusion mode
or a pure visco-elastic mode. The assumption of a single relaxation time for
the entanglements could be an over-simplification \cite{BG77Brochard83}.
Indeed, in the range of strong coupling the observed modes are broadened.
Nevertheless, it is\ clearly seen that far above the critical temperature
the fast mode is closely associated with the diffusion and the slow mode
with the entanglements but that near the critical point the amplitude of the
fast mode vanishes and the slow mode becomes critical. However, the position
of the slow mode is shifted to larger times with respect to the uncoupled
critical diffusion as $\tau _{-}=\tau _{q}\left( 1+q^{2}\xi _{{\rm ve}%
}^{2}\right) $, whereas the position of the fast mode is shifted to shorter
times with respect to the uncoupled visco-elastic time as $\tau _{+}=\tau _{%
{\rm ve}}\left( 1+q^{2}\xi _{{\rm ve}}^{2}\right) ^{-1}$.

In Fig. 4, we have plotted the visco-elastic time $\tau _{{\rm ve}}$, the
visco-elastic length $\xi _{{\rm ve}}$, and the viscosity amplitude $\eta _{%
{\rm 0}}$ (normalized by the solvent viscosity $\eta _{{\rm s}}$)\ as
functions of the molecular weight. We have found that these parameters
apparently diverge along the critical line in the limit of infinite
molecular weight ({\it theta}-point limit) as approximately $M_{{\rm w}%
}^{1.3}$, $M_{{\rm w}}^{0.5}$, and $M_{{\rm w}}^{0.6}$, respectively. We
note that $\tau _{{\rm ve}}$\ scales with $M_{{\rm w}}$ weaker than the
theoretical prediction, $M_{{\rm w}}^{9/4}$, for the ``disentanglement
time'' in {\it theta} solvents at overlap concentrations \cite
{BG77Brochard83}, while $\xi _{{\rm ve}}$\ scales as the radius of gyration $%
R_{{\rm g}}\sim M_{{\rm w}}^{0.5}$ \cite{deGennes,Meln}. Experimental
viscosity data are available only for $M_{{\rm w}}=1.96\times 10^{5}$ \cite
{LaoChu75}. The value\ of $\eta _{{\rm 0}}(T_{{\rm c}})$ obtained from our
treatment of the light-scattering data is lower than that implied by the
viscosity data.\ This fact, also reported in ref. \cite{Jacob:01}, requires
further investigation.

\begin{figure*}
\centering
%\hspace{10pt}
\epsfbox{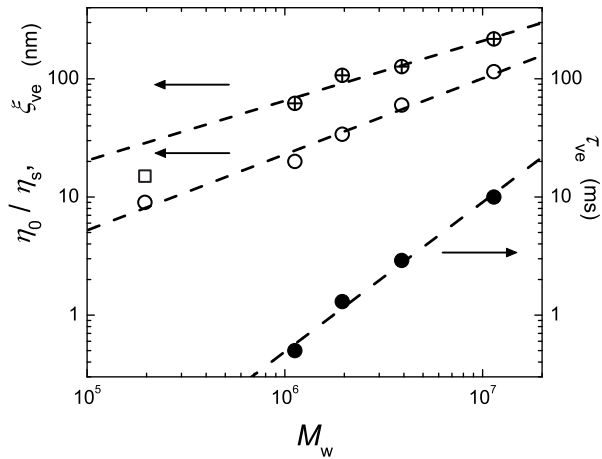}
\vspace{0.4cm}
\begin{minipage}[]{8.5cm}
\protect\caption{
Visco-elastic characteristic length $\xi _{{\rm ve}}$ (crossed
circles, left axis), the background viscosity coefficient $\eta _{{\rm 0}}$
normalized by the solvent viscosity $\eta _{{\rm s}}$ (open circles, left
axis; square from measurements of Lao {\it et al}. \protect\cite{LaoChu75}), and the
relaxation time $\tau _{{\rm ve}}$ (solid circles, right axis) as functions
of the molecular weight $M_{{\rm w}}$. The dashed lines show the slopes of
0.51, 0.64, and 1.3, respectively.
}
\end{minipage}
\label{f:1}
\end{figure*}

The phenomenon of coupling of soft modes has also been observed in polymer
blends \cite{YajimaDouglas} and in sheared polymer solutions \cite{Pine92}.
In near-critical gas-liquid binary fluids, coupling of two diffusion
relaxation modes associated with mass diffusivity and thermal diffusivity
produce two effective dynamic modes, one of them becoming a critical mode 
\cite{Anis2exp}. The difference with polymer solutions is that in simple
fluid mixtures the two original uncoupled modes belong to the same dynamic
universality class. A more straightforward analog of the dynamic coupling in
polymer solutions near the {\it theta} point is expected to appear in $^{3}$%
He--$^{4}$He mixture near its tricritical point.

We are indebted to I. K. Yudin for his help in setting up the
light-scattering instrumentation. We acknowledge valuable discussions with
H. Z. Cummins, J. F. Douglas and R. W. Gammon. The research is supported by
the Division of Chemical Sciences, Geosciences, and Biosciences, Office of
Basic\ Energy Sciences, Department of Energy under Grant No.
DE-FG02-95ER-14509.

%\newpage

\end{multicols}

\vfill

\vspace{-0.4cm}

\begin{references}
\vspace{-0.4cm}

\bibitem{MelnAnis}  Y. B. Melnichenko, M. A. Anisimov, A. A. Povodyrev, G.
D. Wignall, J. V. Sengers, and, W. A. Van Hook, Phys. Rev. Lett. {\bf 79},
5266 (1997).

\bibitem{PREinPress}  M. A. Anisimov, A. F. Kostko, and J. V. Sengers, Phys.
Rev E {\bf 65}, 051805 (2002), cond-mat/0201222.

\bibitem{deGennes}  P. G. de Gennes, {\it Scaling Concepts in Polymer
Physics }(Cornell University, Ithaca, NY, 1979).

\bibitem{Domb}  I. D. Lawrie and S. Sarbach, in: {\it Phase Transitions and
Critical Phenomena}, edited by C. Domb and J. L. Lebowitz, (Academic Press,
New York, 1984), Vol. 9. p.~1.

\bibitem{Tanaka}  H. Tanaka, Y. Nakanishi, and N. Takubo, Phys. Rev. E {\bf %
65}, 021802 (2002).

\bibitem{BG77Brochard83}  F. Brochard and P. G. de Gennes, Macromolecules 
{\bf 10}, 1157 (1977); F. Brochard, J. Physique {\bf 44}, 39 (1983).

\bibitem{AdamDelsanti}  M. Adam and M. Delsanti, Macromolecules {\bf 18},
1760 (1985).

\bibitem{JianBrown96}  T. Jian, D. Vlassopoulos, G Fytas, T. Pakula, and W.
Brown, Colloid Polym. Sci. {\bf 274}, 1033 (1996).

\bibitem{NicolaiStepanek}  T. Nicolai, W. Brown, R. Johnsen, and P.
Stepanek, Macromolecules {\bf 23}, 1165 (1990).

\bibitem{RitzlBelkoura}  A. Ritzl, L. Belkoura, and D. Woermann, Phys.
Chem.-Chem. Phys. {\bf 1}, 1947\ (1999).

\bibitem{Jacob:01}  J. Jacob, M. A. Anisimov, J. V. Sengers, V. Dechabo, I.
K. Yudin, and R. W. Gammon, Appl. Opt. {\bf 40}, 4160 (2001).

\bibitem{PhotocorSP}  I. K. Yudin, G. L. Nikolaenko, V. I. Kosov, V. A.
Agayan, M. A. Anisimov, and J. V. Sengers, Int. J. Thermophys. {\bf 18},
1237 (1997).

\bibitem{Meln}  Yu. B. Melnichenko and G. D. Wignall, Phys. Rev. Lett. {\bf %
78}, 686 (1997);

\bibitem{Malte}  J. M. Schr\"{o}der, S. Wiegand, L. B. Aberle, M. Kleemeier,
and W. Schr\"{o}er, Phys. Chem.-Chem. Phys. {\bf 1}, 3287 (1999).

\bibitem{Ferrell}  R. A. Ferrell, Phys. Rev. {\bf 169}, 199 (1968).

\bibitem{DoiOnuki}  M. Doi and A. Onuki, J. Phys. (Paris) {\bf 2}, 1631
(1992).

\bibitem{Kawasaki}  K. Kawasaki, in {\it Phase Transitions and Critical
Phenomena}, edited by C. Domb and M. S. Green (Academic, New York, 1976),
Vol. 5a, p. 165.

\bibitem{Burstyn}  H. C. Burstyn, J. V. Sengers, J. K. Bhattacharjee, and R.
A. Ferrell, Phys. Rev. A {\bf 28}, 1567 (1983).

\bibitem{Kiselev}  S. B. Kiselev and V. D. Kulikov, Int. J. Thermophys. {\bf %
15}, 283 (1994).

\bibitem{LuettmerSengers}  J. Luettmer-Strathmann and J. V. Sengers, J.
Chem. Phys. {\bf 104}, 3026 (1996); {\bf 106}, 438 (1997).

\bibitem{LaoChu75}  Q. H. Lao, B. Chu, and N. Kuwahara, J. Chem. Phys. {\bf %
62}, 2039 (1975).

\bibitem{BrochdeGenn82}  F. Brochard and P.G. de Gennes, Physicochem.
Hydrodyn. {\bf 4}, 313 (1983).

\bibitem{YajimaDouglas}  H Yajima, D. W. Hair, A. I. Nakatani, J. F.
Douglas, and C. C. Han, Phys. Rev. B 47, 12268 (1993).

\bibitem{Pine92}  P. K. Dixon, D. J. Pine, and X.-l. Wu, Phys. Rev Lett. 
{\bf 68}, 2239 (1992).

\bibitem{Anis2exp}  M. A. Anisimov, V. A. Agayan, A. A. Povodyrev, J. V.
Sengers, and E. E. Gorodetskii,\ Phys. Rev. E {\bf 57}, 1946 (1998).
\end{references}
\end{document}